\documentclass{aa}
\def\folio{\ifnum\pageno=1\nopagenumbers\else\number\pageno\fi}

\def\lax    {\ifmmode{_<\atop^{\sim}}\else{${_<\atop^{\sim}}$}\fi}
\def\gax    {\ifmmode{_>\atop^{\sim}}\else{${_>\atop^{\sim}}$}\fi}
\newbox\grsign      \setbox\grsign=\hbox{$>$} 
\newdimen\grdimen   \grdimen=\ht\grsign
\newbox\simgreatbox \setbox\simgreatbox=\hbox{\raise.5ex\hbox{$>$}\llap
                        {\lower.5ex\hbox{$\sim$}}}\ht1=\grdimen\dp1=0pt
\newbox\simlessbox  \setbox\simlessbox =\hbox{\raise.5ex\hbox{$<$}\llap
                        {\lower.5ex\hbox{$\sim$}}}\ht2=\grdimen\dp2=0pt

\def\d {\phantom{$0$}}
\def\dd {\phantom{$00$}}

\newbox\grsign \setbox\grsign=\hbox{$>$} \newdimen\grdimen \grdimen=\ht\grsign
\newbox\laxbox \newbox\gaxbox
\setbox\gaxbox=\hbox{\raise.5ex\hbox{$>$}\llap
     {\lower.5ex\hbox{$\sim$}}}\ht1=\grdimen\dp1=0pt
\setbox\laxbox=\hbox{\raise.5ex\hbox{$<$}\llap
     {\lower.5ex\hbox{$\sim$}}}\ht2=\grdimen\dp2=0pt
\def\gax{\mathrel{\copy\gaxbox}}
\def\lax{\mathrel{\copy\laxbox}}
\def\boxit#1    {\vbox{\hrule\hbox{\vrule\kern3pt
                  \vbox{\kern3pt#1\kern3pt}\kern3pt\vrule}\hrule}}
\def\h      {\ifmmode{^{\rm h}}\else{$^{\rm h}$}\fi}
\def\m      {\ifmmode{^{\rm m}}\else{$^{\rm m}$}\fi}
\def\s      {\ifmmode{^{\rm s}}\else{$^{\rm s}$}\fi}
\def\decas    {\ifmmode{{\rlap.}{''}}\else{${\rlap.}{''}$}\fi}
\def\mum     {\ifmmode{\mu{\rm m}}\else{$\mu{\rm m}$}\fi}
\def\s      {\ifmmode{^{\rm s}}\else{$^{\rm s}$}\fi}
\def\deg      {\ifmmode{^{\circ}}\else{$^{\circ}$}\fi}
\def\as     {\ifmmode {\rlap.}$\,$''$\,$\! \else ${\rlap.}$\,$''$\,$\!$\fi}
\def\decsec  {\ifmmode {\rlap.}$\,$^{s}$\,$\! \else ${\rlap.}$\,$^{s}$\,$\!$\fi}\def\decs  {\ifmmode {\rlap.}$\,$^{s}$\,$\! \else ${\rlap.}$\,$^{s}$\,$\!$\fi}

\def\kms    {\ifmmode{{\rm km~s}^{-1}}\else{km~s$^{-1}$}\fi}

\def\Lsun   {$L_{\odot}$}

\def\Mspy   {\ifmmode {M_{\odot} {\rm yr}^{-1}} \else $M_{\odot}$~yr$^{-1}$\fi}
\def\Mdot   {\ifmmode {\dot M} \else $\dot M$\fi}
\def\mhd    {\ifmmode {n_{{\rm H}_2}} \else $n_{{\rm H}_2}$\fi}
\def\mhcd   {\ifmmode {N_{{\rm H}_2}} \else $N_{{\rm H}_2}$\fi}

\def\El      {\ifmmode{E_{\ell}}\else{$E_{\ell}$}\fi}
\def\beam    {\ifmmode{\theta_{\rm B}}\else{$\theta_{\rm B}$}\fi}
\def\mjyb   {\ifmmode {{\rm mJy~beam}^{-1}} \else{mJy~beam$^{-1}$}\fi}
\def\mujyb   {\ifmmode {\mu{\rm Jy~beam}^{-1}} \else{$\mu$Jy~beam$^{-1}$}\fi}
\def\Trot   {\ifmmode{T_{\rm rot}}\else$T_{\rm rot}$\fi}    
    
\def\Teff   {\ifmmode{T_{\rm eff}}\else$T_{\rm eff}$\fi}

\def\TMB    {$T_{\rm MB}$}

\def\ITRS   {\ifmmode{\smallint {\rm T}_{R}^{*}dv}\else{$\smallint 
{\rm T}_{R}^{*}dv$}\fi}
\def\ITRS   {\ifmmode{\smallint {\rm T}_{R}^{*}dv}\else{$\smallint 
{\rm T}_{R}^{*}dv$}\fi}
\def\ITAS   {\ifmmode{\smallint {\rm T}_{A}^{*}dv}\else{$\smallint 
{\rm T}_{A}^{*}dv$}\fi}

\def\hh         {H$_2$}

\def\hzo        {H$_2$O}

\def\nuteo {$\nu_2$=1}
\def\nuteo {\mbox{$\nu_2$=1}}

\def\vwo   {\hbox{$5_{50}-6_{43}$}}
\def\vwt   {\hbox{$4_{40}-5_{33}$}}

\def\ogs   {\hbox{$1_{10}-1_{01}$}}
\def\tnt   {\hbox{$6_{61}-7_{52}$}}
\def\tns   {\hbox{$6_{60}-7_{53}$}}
\def\tts   {\hbox{$5_{23}-6_{16}$}}
\def\lefttitle#1  {\noindent \hangindent=18.0pt \hangafter=1 {#1} \par}
\def\vol#1  {{\bf {#1}{\rm,}\ }}
\font\tenssb=cmssbx10
\font\tenbf=cmbx10
\font\sevenbf=cmbx8
\font\fivebf=cmbx6
\def\unetdemi    {\smallskipamount=6pt plus2pt minus2pt
                  \medskipamount=12pt plus4pt minus4pt
                  \bigskipamount=24pt plus8pt minus8pt
                  \normalbaselineskip=16pt plus0pt minus0pt
                  \normallineskip=2pt
                  \normallineskiplimit=0pt
                  \jot=6pt
                  {\def\smallskip {\vskip\smallskipamount}}
                  {\def\medskip   {\vskip\medskipamount}}
                  {\def\bigskip   {\vskip\bigskipamount}}
                  {\setbox\strutbox=\hbox{\vrule 
                    height17.0pt depth7.0pt width 0pt}}
                  \parskip 12.0pt
                  \normalbaselines}
\def\smallerspace {\smallskipamount=3pt plus0pt minus0pt
                  \medskipamount=6pt plus0pt minus0pt
                  \bigskipamount=10.5pt plus0pt minus0pt
                  \normalbaselineskip=10.5pt plus0pt minus0pt
                  \normallineskip=1pt
                  \normallineskiplimit=0pt
                  \jot=3pt
                  {\def\smallskip {\vskip\smallskipamount}}
                  {\def\medskip   {\vskip\medskipamount}}
                  {\def\bigskip   {\vskip\bigskipamount}}
                  {\setbox\strutbox=\hbox{\vrule 
                    height8.5pt depth3.5pt width 0pt}}
                  \parskip 0pt
                  \normalbaselines}
\def\memospace    {\smallskipamount=4pt plus1pt minus1pt
                  \medskipamount=6pt plus2pt minus2pt
                  \bigskipamount=14pt plus6pt minus6pt
                  \normalbaselineskip=14pt plus0pt minus0pt
                  \normallineskip=1pt
                  \normallineskiplimit=0pt
                  \jot=4pt
                  {\def\smallskip {\vskip\smallskipamount}}
                  {\def\medskip   {\vskip\medskipamount}}
                  {\def\bigskip   {\vskip\bigskipamount}}
                  {\setbox\strutbox=\hbox{\vrule 
                    height17.0pt depth7.0pt width 0pt}}
                  \parskip 2.0pt
                  \normalbaselines}
\def\memowidespace    {\smallskipamount=5pt plus1pt minus1pt
                  \medskipamount=7.5pt plus2pt minus2pt
                  \bigskipamount=17.5pt plus6pt minus6pt
                  \normalbaselineskip=17.0pt plus0pt minus0pt
                  \normallineskip=1.25pt
                  \normallineskiplimit=0pt
                  \jot=5pt
                  {\def\smallskip {\vskip\smallskipamount}}
                  {\def\medskip   {\vskip\medskipamount}}
                  {\def\bigskip   {\vskip\bigskipamount}}
                  {\setbox\strutbox=\hbox{\vrule 
                    height21.25pt depth8.75pt width 0pt}}
                  \parskip 2.5pt
                  \normalbaselines}
\usepackage{graphicx}
\usepackage{natbib}
\usepackage{txfonts}
\usepackage{ulem}
      \def\new#1 {{\bf #1 }}
      \def\cut#1 {\sout{#1} }

\begin{document}

\title{Submillimeter vibrationally excited 
water emission from the peculiar red supergiant VY CMa}
\author{K. M. Menten
\inst{1}
\and
S. D. Philipp
\inst{1}
\and
R. G\"usten
\inst{1}
\and
J. Alcolea
\inst{2}
\and
E. T. Polehampton
\inst{3}
\and
S. Br\"unken
\inst{4}
}


\offprints{K. M. Menten}

\institute{Max-Planck-Institut f\"ur Radioastronomie,
Auf dem H\"ugel 69, D-53121 Bonn, Germany
\email{kmenten, sphilipp, rguesten@mpifr-bonn.mpg.de}
\and
Observatorio Astronomico Nacional,
C/ Alfonso XII, No 3, E-28014 Madrid, Spain
\email{j.alcolea@oan.es}
\and
Space Science and Technology Department Rutherford Appleton Laboratory,
Chilton, Didcot,
Oxfordshire, OX11 0QX, UK
\email{E.T.Polehampton@rl.ac.uk}
\and
Harvard-Smithsonian Center for Astrophysics,
60 Garden Street,
Cambridge, MA 02138, USA
\email{sbruenken@cfa.harvard.edu}
}

\date{Received / Accepted}
\titlerunning{Vibrationally excited \hzo\ in VY CMa}
\authorrunning{Menten et al.}

 \abstract
   {Vibrationally excited emission from the SiO and \hzo\ molecules probes the
   innermost circumstellar envelopes of oxygen-rich red giant and supergiant stars.
   VY CMa is the most prolific known emission source in these molecules.}
   {Observations were made to search for 
   rotational lines in the
   lowest vibrationally excited state of \hzo.} 
   {The APEX telescope was used for observations of \hzo\ lines at frequencies around
   300 GHz.}
   {Two vibrationally excited  \hzo\ lines were detected, 
   a third one
   could not be found.
   In one of the lines we find evidence for weak maser action,
   similar to known (sub)millimeter \nuteo\ lines. 
   We find that the other line's intensity 
   is consistent with thermal excitation by the circumstellar infrared radiation field.
   Several SiO lines were detected
   together with the \hzo\ lines.}

\keywords{Stars: AGB and post-AGB  -- Stars: individual: VY CMa -- supergiants -- circumstellar matter}

\maketitle

\section{\label{intro}Introduction}

In the past, millimeter-wavelength rotational lines from the lowest
vibrationally excited state of water (H$_2$O), the \nuteo\ bending mode,
have been detected from the innermost envelopes of oxygen-rich
giant and supergiant stars.
\citet{MentenMelnick1989} used the IRAM\,30m telescope to discover the \nuteo,
 $J_{K_{\rm a},K_{\rm c}}$=\vwo\ line near 232\,GHz toward the high mass-loss supergiant VY\,CMa
and the nearby semiregular variable W\,Hya. Toward the former they also
detected the 96\,GHz \vwt\ transition.
Negative results were obtained for a number of other stars.
All of the observed spectra contained a narrow (width $\approx 1$\,\kms)
feature, clearly indicating maser action.
This feature also shows up in the spectrum
of the
658\,GHz \nuteo, \ogs\ ortho rotational ground-state transition, which has
been found to show very strong maser action not only in VY\,CMa and W\,Hya, but in a
variety of other (super)giant stars \citep{MentenYoung1995}. The excitation of these
masers is discussed in Sect. \ref{vibexcitation}.

There is also the interesting possibility that
{\it thermal} emission from hot water has been detected:
The 232\,GHz spectra of, both, VY\,CMa and W\,Hya
show, in addition to the maser spike, broad emission over the velocity range covered by
the circumstellar outflows of these objects.

The target of the observations 
is the peculiar red supergiant VY\,Canis Majoris.
With a luminosity of $3.5\,10^5$\,\Lsun\ \citep{Sopka1985}
and a mass-loss rate of  2--$3\,10^{-4}$\,\Mspy\ \citep{Danchi1994}
VY\,CMa is a remarkable object by any standard. For our discussions
it is relevant to note that a plethora of maser lines from (ground-state and \nuteo) water
and all the Si isotopes of silicon monoxide (from the v = 0 to 5 states)
have been detected from this object. At far-infrared wavelengths
its spectrum is completely dominated by \hzo\ lines including several lines from
the \nuteo\ state \citep{Neufeld1999}.

We used APEX to search for the
\hzo\ \nuteo,
\tnt, \tns, and \tts\ lines near 294, 297, and 336\,GHz.
Our observations 
are summarized in Sect. \ref{obs}.
In Sect. \ref{results}
we report the  detection of  the first and
the third, but not of the second transition and present the observed spectra  along
with spectra of simultaneously observed silicon monoxide lines.
After summarizing
(in Sect.\,\ref{vibexcitation})
our phenomenological picture of vibrationally excited
\hzo\ maser excitation, in Sect.\,\ref{nature} we investigate the
possibility whether either or both of the detected lines
may be the result of thermal excitation
and come to the conclusion that this could be
the case for one of them.



\section{\label{obs}APEX observations and data reduction}
Our observations were made in 2005 July/August under generally
excellent weather conditions
with the 12\,m Atacama Pathfinder Experiment
telescope (APEX\footnote{This publication is based on data
  acquired with the Atacama Pathfinder Experiment (APEX). APEX is a
  collaboration between the Max-Planck-Institut f\"ur Radioastronomie,
  the European Southern Observatory, and the Onsala Space
  Observatory.}; see G\"usten et al.; this issue).

The APEX 2a facility receiver
was used (Risacher et al., this issue), which covers the 278--375\,GHz frequency range that
contains the frequencies of several \hzo\ lines including the ones we observed.
Two isotopic silicon monoxide lines
were in the same band as the 336.2\,GHz line.
A number of other lines were also observed including the $^{29}$SiO
v=0, $J$=8$-$7 line.
All lines considered here are listed in Table\,\ref{lines}
and are discussed in Sect.\,\ref{results}.
Calibration was obtained using the standard
chopper wheel technique. The radiation was analyzed with the MPIfR
Fast Fourier Transform spectrometer that provides 16384 frequency
channels (Klein et al., this issue) over the 1\,GHz intermediate frequency bandwidth.
To increase the signal to noise ratio, the spectra were
smoothed to effective velocity resolutions appropriate 
for the measured linewidths, typically $~1$ \kms. To check the telescope pointing
the receiver was tuned to the 345.8\,GHz CO $J=$3$-$2   line.
We measured 5 point rasters in azimuth and elevation centered on VY CMa itself
and pointing corrections were determined by least squares fits.
The pointing
was found to be accurate to within $\sim4''$, acceptable
given the
beam size, $\theta_{\rm B}$, which is $18''$ FWHM at 345.8 GHz.
We present our line intensities in a main-beam brightness temperature
($T_{\rm MB}$) scale and assumed a main-beam efficiency, $\eta_{\rm MB}$,
equal to the observationally
determined value at 345 GHz, i.e. 0.70.
We estimate that our intensity scale is accurate to within 15\%.
The FWHM beam size, $\theta_{\rm B}$, is $19''$ at 336\,GHz and $21''$ at
294\,GHz. 
\begin{table*}[tb]
\begin{center}
\caption{\label{lines}Water and silicon monoxide lines observed with APEX.}
\begin{tabular}{llllclll}
 \hline \hline
\hzo~\nuteo       &Frequency     &$E_\ell$&$\int$\TMB~dv           & v-range      & $\Delta$v & \TMB & v$_{\rm LSR}$ \\
$J_{K_{a}K_{c}} =$ & (MHz)              &   (K)  & (K km~s$^{-1})$ & (\kms)       & (\kms) & (K)& (\kms)\\
$6_{61} - 7_{52}$ &  293664.4$^{\rm a}$&   3920 &  \d1.3(0.1)      & $\sim[11,32]$& 22(3)      & 0.055   & $\sim21$\\
$6_{60} - 7_{53}$ &  297439.1$^{\rm a}$&   3920 &  $<0.16^{\rm b}$&   --          &  \d\d --   & $<0.029^{\rm c}$\\
$5_{23} - 6_{16}$ &  336227.6$^{\rm a}$&   2939 & \d1.2(0.1)       & $\sim[11,32]$& 18 (2)     & 0.065   &$\sim21$ \\
$J=8-7$\\
$^{29}$SiO~v$=3$   &  335880.3$^{\rm d}$    &   5280 & \d0.78(0.07) & [14,32]      &\d8.4(0.9)        & 0.087& 23.5(0.3)\\
$^{30}$SiO~v$=1$   &  336602.4$^{\rm d}$  &   1804 & 41.3(0.5)      & \d[-9,64]    &$\sim 2^{\rm e}$& $3.1^{\rm e}$& $19.5^{\rm e}$ \\
$^{29}$SiO~v$=0$   &  342980.6$^{\rm d}$  &  \dd58 & 34.2(0.3)      & [$-16,62$]   &39.6(0.5)       & 0.81 & 21.9(0.2) \\
\noalign{\smallskip}
 \hline
 \noalign{\smallskip}
 \end{tabular}
\end{center}

Columns are (from left to right) quantum numbers of upper and lower state,
frequency, energy above ground of lower state, integrated main-beam brightness temperature,
velocity range covered by line (FWZP), FWHM line width from a Gaussian fit,
peak main-beam brightness temperature, and centroid LSR velocity,
Multiply  \TMB(K) by 30 to get flux density in Jy. Numbers in parentheses are formal errors derived
from Gaussian fitting and/or noise level analysis.
$^{\rm a}$\citet{Chen2000}
$^{\rm b}$$3\sigma$ upper limit assuming a 20 \kms\ wide velocity range.
$^{\rm c}$$3\sigma$ upper limit (in a spectrum smoothed to 1.5 km~s$^{-1}$ resolution).
$^{\rm d}$Calculated from high precision molecular constants measured by \citet{Molli1991}.
$^{\rm e}$Value for strongest velocity component.

\end{table*}

\section{\label{results}Results}
\begin{figure}[h]
\begin{center}
\includegraphics[width=6.5cm,angle=0]{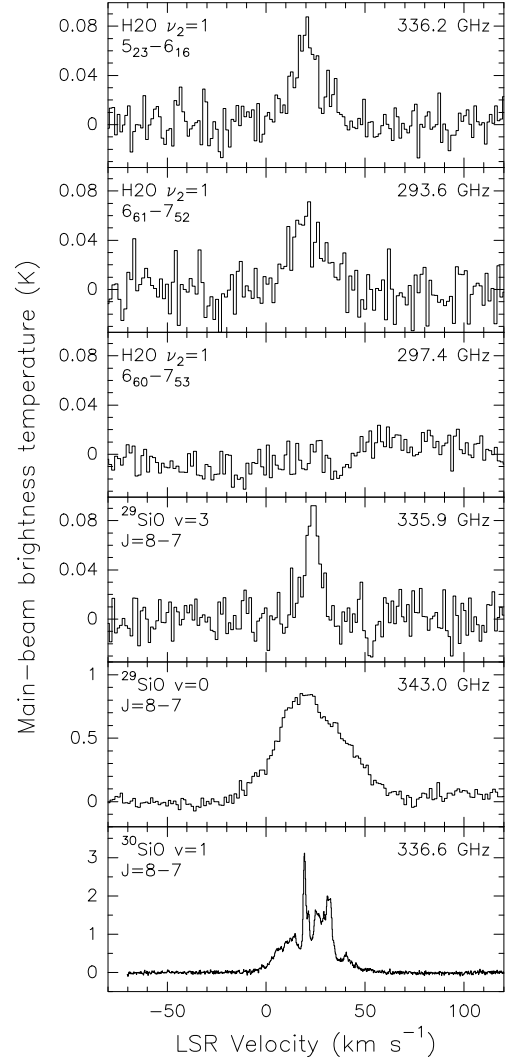}
\caption{\label{observedspectra}\textit{Top to bottom:} Spectra
observed toward VY CMA of the \hzo\
\nuteo, \tts, \tnt, and \tns\ lines and the $J$=8$-$7 lines of $^{29}$SiO
in the  v=3 and
1 states and $^{30}$SiO in the v=1 state. The spectra have been smoothed to
channel spacings of 1.3, 1.5, 1.5, 1.3, 1.3, and 0.22\,\kms, respectively.}
\end{center}
\end{figure}

Fig. \ref{observedspectra} presents our
observed spectra; see also Table \ref{lines}. We detect the two
ortho-\hzo\ lines (\tts\ and \tnt), but not the latter line's
para \hzo-equivalent (\tns).
Given that the 293 and 297\,GHz lines 
Einstein A-values and
energies above ground are very similar, we would expect an intensity ratio
of $\approx$\,3:1 for these lines if they were thermally excited.
For the
integrated intensities we obtain a lower limit of 8.2 for the ratio of the ortho to the para line,
suggesting weak maser action
that boosts the ortho line's intensity more than the para line's.
The silicon monoxide lines are all
between the $J$=8 and 7 rotational levels but within
different vibrational states of vastly different energy above ground state.
The $^{30}$SiO, v=1 line clearly shows maser action expressed in multiple
emission spikes while all the other lines' profiles are consistent
with (quasi)thermal emission.
All lines are centered
around 20\,\kms\ which is within a few km~s$^{-1}$ of the stellar velocity
(whose exact value has an uncertainty of a few \kms). The velocity
range covered by the $^{29}$SiO, v=0 line is widest and is comparable to
that observed in other thermally excited lines of SiO and other
molecules (e.g. \citet{Nyman1992}). 
The v=3\,$^{29}$SiO line and the two detected \hzo\ lines are reasonably
well fit by Gaussians.  The very narrow width of v=3\,$^{29}$SiO line ($\sim 8$\,\kms)
indicates maser action. The \hzo\ lines have about half the width of the
$^{29}$SiO, v=0 line (and much poorer signal-to-noise ratio) and their
shapes do not readily suggest masing, but, possibly, that they are formed in the
innermost envelope where the stellar outflow has not yet reached its terminal
velocity.

\section{\label{disc}Discussion}
\subsection{\label{vibexcitation}Vibrationally excited water masers}

\citet{AlcoleaMenten1993} noted that the \nuteo\ maser lines
described above seem to result from a systematic overpopulation of
(what they call) the ``transposed backbone'' levels of the energy level diagram, (see their Fig. 1),
 i.e.
levels with quantum numbers $K_{\rm a}$=$J$ and  $K_{\rm c}$=0 and 1 for para- and ortho-\hzo,
respectively, with respect
to their neighboring levels.
This situation  is just opposite from that
found in the ground state, where many of the  ``backbone'' levels
($K_{\rm a}$=0 or 1  and  $K_{\rm c}$=$J$)
may become overpopulated.
This systematic inversion may be explainable by means of differential
trapping in the radiative vibrational decays of the levels involved
in those maser transitions and, possibly, by pumping  to and decay from
higher vibrational states. Detailed calculations to support this
are still extant and not yet feasible, since collisional rates
for vibrational transitions are unknown and can only be guessed (see Sect.\,\ref{nature}).

The mechanism described above leads
to the excitation of the known  \vwt, \vwo, and \ogs\
masers at 96, 232, and 658\,GHz, respectively, and
predicts inversion also, a.o., of the 294 and 297\,GHz \tnt\ and
\tns\ lines, of which we only detect the former one. However, given the
uncertainties in our understanding of the maser excitation process it is by no means
clear that these lines should be inverted at all.
Certainly, there is no  a priori  reason to expect maser emission in the 336\,GHz line.

Given the spectral similarities between SiO maser lines and \nuteo\
\hzo\ lines pointed out by \citet{MentenMelnick1991} it is highly likely
that  \nuteo\ \hzo\ \textit{maser} emission arises from the same region as
the multi-isotope/multi-v/multi-$J$ SiO maser emission modeled by
\citet{Gonzalez1997}\footnote{In particular, see  Fig. 4 of \cite{MentenMelnick1991},
which shows that single  tell-tale
narrow ($\sim 1$
\kms -wide), long lived ($> 10$\,y) feature at 22.3 \kms\ is seen in
SiO maser lines from all three Si isotopes, as well as in both \nuteo\
lines. In fact, strikingly, the \vwt\ \hzo\ and  the $^{29}$SiO and $^{30}$SiO
v=0, $J$=1$-$0 lines
show \textit{only} emission in this feature.}.
%

It may thus be expected that for some of the known \nuteo\ lines part or all the emission is masing,
although the excitation mechanism is still uncertain and may involve IR line overlaps.
Could we, on the other hand, expect to observe thermal emission from a region of a size
similar to that of the SiO maser region in lines unaffected by any inversion mechanism?

\subsection{\label{nature}The  nature of the water lines detected with APEX}
To investigate whether the observed emission in one or both of the
\hzo\ lines detected by us might be of thermal nature, we make the following considerations.
\textit{Assume} that the line's level populations were in Local Thermodynamic
Equilibrium (LTE), meaning
$T_{\rm ex}=T_{\rm rot}=T_{\rm vib}=T_{\rm kin}$, i.e., that
excitation, rotational, and vibrational temperatures
are all equal to the kinetic temperature in the emitting region
and that the lines were thermalized, i.e. their level populations
were described by a Boltzmann distribution at that temperature. If we assume a uniform medium,  the
line's main-beam brightness temperature, $T_{\rm MB}$, is
given just by $f \times T_{\rm kin}(1 - e^{-\tau})$, where the beam-filling factor, $f$,
is given by $(\theta_{\rm S}/\theta_{\rm B})^2$.  $\theta_{\rm S}$ and
$\theta_{\rm B}$ are the FWHM size of the Gaussian source and beam.

We now estimate whether plausible assumptions on the emission region
could produce optically thick emission.

Around VY CMa, vibrationally (v=1 and 2) SiO masers arise from a region
of radius $\sim 0\as15$ (Menten \&\ Reid, in preparation).
Assuming a distance of 1.5 kpc for VY CMa \citep{LadaReid1978}, this corresponds to
$3\,10^{15}$\,cm or 20 stellar radii. Inside that region, SiO pumping models
require a molecular hydrogen density, $n({\rm H}_2)$, between $10^9$ and
$10^{10}$\,cm$^{-3}$ \citep{LockettElitzur1992, Bujarrabal1994}.

In the following we assume a constant density of $2~10^9$\,cm$^{-3}$ within $10^{15}$\,cm.
This seems to be a plausible average value between  $n({\rm H}_2)$\,=\,$7\,10^7$\,cm$^{-3}$, the value
\citet{Danchi1994} find for the  inner radius of the dust shell and the density  of
$10^{12}$\,cm$^{-3}$ in the stellar photosphere; see, e.g., \citet{reid1997}.
Danchi et al. derive  $T$\,=\,1563\,K at the inner dust shell radius, while
\citet{LeSidaner1996} and \citet{Massey2006} derive a temperature of
750 -- 850\,K for the dust from IR photometry. In the following we adopt 1000\,K
for the temperature.
Assuming a homogeneous medium,  we obtain an  column density of
$2~10^{24}$\,cm$^{-2}$.

\textit{If} we assume that the observed \nuteo\ lines are thermalized we
now can calculate their optical depths.
For this we have used  A-values and level energies from the HITRAN
database \citep{Rothman2003} and a  value for the partition function,
$Q(1000~{\rm K})$,  of 1194
from \citet{Chen2000}. We use the upper range of the [\hzo/\hh]
abundance ratios of $4~10^{-4}$
that \citet{GonzalezCernicharo1999} derive for stars with high
mass-loss rates while noting that
\citet{Zubko2004} derive $4\,10^{-4}$ for VY\,CMa specifically
in a completely
independent analysis.

For $T$\,=\,1000\,K
we compute  optical depths of 1.0, 0.36, and 2.8 for the  
294, 297, and 336\,GHz lines, respectively, and  \TMB-values of 12, 5, and 28\,mK.
The lower value of the 297\,GHz para-line as compared to the
294\,GHz ortho-line is due to the 3:1 ortho-to-para ratio.

The calculated value for the 336\,GHz lines is only a factor of 2.3 lower than the
actually measured intensity, while that ratio for the 293 GHz line
is larger, $\sim5$.
For the broad component of the 232.6\,GHz \vwo\ line detected by \citet{MentenMelnick1989} with a
$12''$ FWHM beam
we calculate \TMB\,=\,43\,mK for $T$\,=\,1000\,K, which is smaller than the 140 mK they observe.
As discussed in Sect. \ref{vibexcitation} one might expect inversion for both of the latter lines
and the weak maser
action they do indeed seem to show (see Sect. \ref{results}) can explain their intensities.

The above estimates assume that our  \hzo\ lines are thermalized.
Thermalization can either be achieved by collisional excitation or radiative excitation.
To check the potential of collisional excitation, we calculate the lines'
critical densities. The critical density of a transition from level $i$ to level
$j$ is given by  $\sum A_{ij}/\sum \gamma_{ij}$, where the sums are over all possible
transitions originating from level $i$. $A_{ij}$ are the Einstein A-coefficients
(again from HITRAN) and $\gamma_{ij}$
the collisional rate coefficients,
calculated by  \citet{Green1993} for different $T$-values and accessed
online via the  Leiden Atomic and
Molecular Database; \citep{Schoier2005}.
Rate coefficients for collisions from the ground to the \nuteo\
state are unknown. In  the following we assume them to be identical to transitions
within the ground state with identical $J, K_{\rm a}$ and $K_{\rm c}$ quantum numbers
multiplied by a factor $c_{\rm vg}$; the study of \citet{Gonzalez2002} suggests $c_{\rm vg}$\,=\,0.02.
For  all of our three \nuteo\ transition we calculate $\sum A_{ij}$\,=\,21\,s$^{-1}$
within 10\%.
The sums of the collisional rate coefficients  for the three lines
are very similar as well:
we calculate (within 5\%)
$\sum \gamma_{ij}$\,=\,$c_{\rm vg}~2.1\,10^{-10}\,{\rm cm}^3\,{\rm s}^{-1}$
for 1000\,K.
From this we derive $n_{\rm crit}$\,=\,$c^{-1}_{\rm vg}\,10^{11}~{\rm cm}^{-3}$.
Comparing this with the densities discussed above ($\sim 10^8$--$10^{10}$\,cm$^{-3}$)
we find that our lines are very unlikely to be thermalized by collisions in the whole $10^{15}$\,cm radius
region.

As to radiative excitation: According to \citet{CarrollGoldsmith1981},
in a region with a strong IR field the
criterion for radiative excitation to determine the populations of energy levels of a
rotational line that is coupled by IR radiation of frequency $\nu_{\rm IR}$ to
levels in an excited vibrational state is
$f/(e^{h\nu_{\rm IR}/kT_{\rm S}} - 1) > A_{\rm rot}/A_{\rm vib}$. Here $A_{\rm vib}$ and
$A_{\rm rot}$ are the A-values the rotational
and vibrational transitions involved, respectively, and $T_{\rm S}$ is the effective temperature
of the IR source. For our case the filling factor, $f$, is equal to 1
and $\nu_{\rm IR} \sim 4.5~10^4$\,GHz (corresponding to 6.7\,$\mu$m), which results
in the ratio in left hand side of the equation to being $\approx 0.2$. This is much larger
than the right hand side, since  $A_{\rm vib} \sim 20\,{\rm s}^{-1}$ and $A_{\rm rot} \sim
10^{-6}\,{\rm s}^{-1}$. Given this, it seems safe to assume that the level populations of the
submillimeter lines are thermalized at the temperature of the emitting region. 

%

In summary, the level populations of the observed \hzo\ \nuteo\
lines  are likely to be excited via infrared radiation.
Making simple, but plausible assumptions
on their dense, hot emission region, 
we can roughly reproduce the observed intensity of the 336 GHz line
assuming thermal excitation. Known characteristics of \hzo\ \nuteo\ masers
lead one to expect that the 294 and 297 GHz lines might have inverted populations.
Indeed, the observed intensity of the 294 GHz ortho-line suggests
that it might be boosted by weak maser action. The
non-detection of its twin para-\hzo\ 297 GHz line is naturally explained
by the lower abundance of the para species and less amplification.
Future imaging with ALMA, whose highest resolution will resolve
the lines'  emission region, will yield important information
on this higly interesting environment.

\bibliographystyle{aa}
\bibliography{5458}

\clearpage

\end{document}